\documentclass {jaa}
\usepackage[english]{babel} 
\usepackage{graphics}
\usepackage{graphicx}
\usepackage{color}
\usepackage{amsmath}
\usepackage{psfrag}
\usepackage{epsfig}
\usepackage[round]{natbib}

\begin{document}
\title[Receiver for the Ooty Wide Field Array]{The Receiver System for the Ooty Wide Field Array}
\author [Subrahmanya et al.]
{C.R. {Subrahmanya${^1}$}, P. {Prasad${^1}$}, B.S. {Girish${^1}$}, R. {Somashekar${^1}$}, \\
\Large\rm P.K. {Manoharan${^2}$} and A.K. {Mittal${^2}$}\\ \\
    ${^1}$Raman Research Institute, C.V. Raman Avenue, Sadashivnagar, Bengaluru 560080 \\
    ${^2}$Radio Astronomy Centre, NCRA-TIFR, P.O. Box 8, Udhagamandalam (Ooty) 643001 \\
}
\date {}
\maketitle

\begin{abstract}
The legacy Ooty Radio Telescope (ORT) is being reconfigured
as a 264-element synthesis telescope, called the Ooty Wide Field
Array (OWFA). Its antenna elements are the contiguous 1.92 m sections
of the parabolic cylinder.  It will operate in a 38-MHz frequency band
centred at 326.5 MHz and will be equipped with a digital receiver including
a 264-element spectral correlator with a spectral resolution of 48
kHz.  OWFA is designed to retain the benefits of equatorial mount,
continuous 9-hour tracking ability and large collecting area of the
legacy telescope and use modern digital techniques to enhance the
instantaneous field of view by more than an order of magnitude.
OWFA has unique advantages for contemporary investigations related
to large scale structure, transient events and space weather watch.
In this paper, we describe the RF subsystems, digitizers and
fibre optic communication of OWFA and highlight some specific aspects
of the system relevant for the observations planned during the
initial operation.
\end{abstract}

\begin{keywords}{radio astronomy: radio telescope, receiver system - 
space weather watch: solar wind transients - cosmology: large scale 
structure of universe - fast radio bursts}
\end{keywords}

\section{Introduction}    
Ooty Radio Telescope (ORT) is an equatorially mounted cylindrical
radio telescope with a parabolic reflector of size  530 m $\times$ 30 m, which
was commissioned in 1970 as a multi-beam radio telescope to operating
over a bandwidth up to $\sim$10 MHz centred at 326.5 MHz \cite{swarup71}.  It is
equipped with a uniformly spaced array of 1056 half-wave dipoles spanning 506 m
of the focal line.  It is located on a hill whose slope is equal
to the latitude of the place, so
that continuous tracking of a celestial object is enabled by
mechanical rotation around a north-south axis.  Using a centrally
controlled phase shifter installed at each dipole, it is possible
to establish a uniform phase gradient along the feed array to steer
the beam in a desired north-south direction. The last major upgrade
of its front-end was about 25 years ago \citep{selvanayagam93}, when each dipole was
equipped with an LNA and an individually controllable phase shifter.
The legacy receiver system consists of  analog beam formation by
phasing the array to form a set of 12 beams spanning $0.6^{\circ} sec(\delta)$
in the north-south direction.   A brief outline of the front end
system is given below. This sets the context for the ongoing
upgrade.

 \begin{figure}
   \centerline{\includegraphics[width=\linewidth]{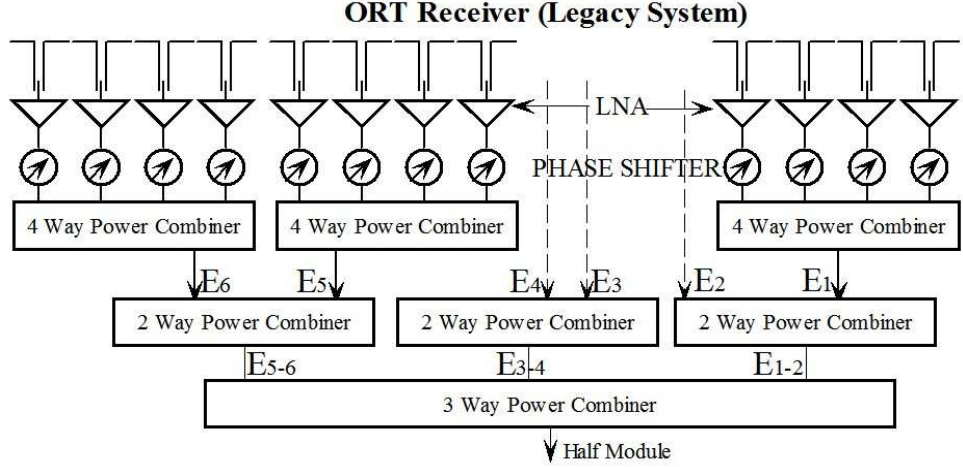}}
   \caption{Legacy Half-Module}
   \label{fig:half-module}
 \end{figure}

The legacy front-end of ORT is organized into 44 identical groups
(“half-modules”) each spanning a 11.5-m section of the focal line (occupied
by 24 consecutive dipoles) centred on a supporting frame. The output
of each dipole is amplified using a low-noise amplifier (LNA) and
passed through a (user-controllable) phase-shifter.  The front-end
electronics of each “half-module” is mechanically housed in identical
enclosures, each catering to a set of 24 consecutive dipoles with
LNA and phase shifter followed by a passive combiner tree as shown
in Figure \ref{fig:half-module}.  The final output of the combiner tree
results in a half-module beam whose phase-centre is typically arranged at the
centre of the “module” constituted by a pair of half-modules on
either side of a supporting frame.  A mixer unit is located on the
supporting frame corresponding to each module, where the two
half-module beams are brought together, combined and down-converted
to form a “module beam” using a centrally distributed local oscillator
at 296.5 MHz.  The resulting module beams (centred at 30 MHz) are
transported to central receiver room using equal length coaxial
cables.  The entire range of receivers used prior to OWFA corresponds
to backend systems for these IF beams. The oldest and the most
widely used receiver is an analog beam forming network for phasing
the 22 modules in 12 different directions in the north-south direction
spanning a total extent of $\sim$$0.6^{\circ} sec(\delta)$. For a summary
of various projects undertaken during the initial decades of operation
of ORT, we refer the reader to \cite{swarup91}. Since then a digital
receiver  was built by \cite{PrabuThesis} for co-location with the
analog beam forming network which allows one to record the digitized
IF beams from all 22 modules and use an off-line processing to enhance
the instantaneous field of view to $\sim$$2.4^{\circ} sec(\delta)$.
Due to the combination of large collecting area, equatorial mount and
the large number of phase-controlled dipoles along the focal line, one
would expect to get a tremendous advantage for contemporary large
scale surveys with the ORT.  But these benefits have been completely offset by
the severe limitation of instantaneous field of view imposed by the
legacy receivers which treated the module beams available in the
receiver room as the primary source of signal.  Hence, we decided
to give a new lease of life to the legacy telescope by carrying out
a major revision of front-end as described in this paper.  In
particular, our efforts have led to a reconfiguration of the telescope
into a synthesis telescope - the Ooty Wide Field Array (OWFA) - in
which the phased sum of 4 consecutive dipoles along the focal line
is treated as an individual antenna. This upgrade enables a range
of different scientific programs including observations of the large
scale distribution of neutral hydrogen at $z \sim 3$, observations of
space weather in the inner heliosphere, and surveys for radio transients
(for more details, refer to Subrahmanya, Manoharan \& Chengalur
(2016, this issue)).

The work related to OWFA began as an in-house activity at the Raman
Research Institute (RRI) where conceptual design and prototyping were
carried out with the help of a couple of small scale industries in
Bengaluru. As part of a feasibility study, a preliminary design was
undertaken for critical subsystems related to data acquisition and
transportation to central computer. The hardware subsystems thus
developed were adequate for working out the logistics for RF digitization,
clock distribution, communication network topology and protocols  as
well as for a configurable correlator.  They were used to re-configure
the ORT as an array of 40 elements where the individual elements were
half-module beams. For a detailed description of this precursor system
(also called the ``Phase I'' system) and some results from initial
tests, we refer the reader to \cite{HiPC2010, PeeyushThesis}.
This system was used as a test bed to work out a suitable communication
network and a practical set of communication protocols useful for the
digital receiver. Experience with this system was used to formulate a
scheme called the Networked Signal Processing System (NSPS,
\cite{NSPS, PeeyushThesis})  which became the basis of an 800-channel
spectral correlator for the 264-element OWFA. This includes a segment
which provides a firmware interface to help realizing the software
correlator on a many-core processor based commodity High Performance
Cluster (HPC) using a Single Program Multiple Data (SPMD) strategy.

In this paper, we give a detailed description of the RF and digitizer
subsystems installed in the field and the fibre-optic communication
system established to transport digitized data to the central receiver
system as well as uplink signals related to time synchronization and
command interface to the digitizers.  Details of the software correlator
will be provided elsewhere.

\section{Ooty Wide Field Array}

Due to limitations of in-house manpower, the detailed design and
fabrication of all subsystems of OWFA were outsourced to small scale
industries in Bengaluru. Augmentation of the infrastructure at the
observatory and local logistics support required for system
commissioning and testing were provided by the Radio Astronomy
Centre, Ooty (Ooty is now known as Udhagamandalam). In order to minimize 
the telescope downtime
and for efficient testing of new systems, the entire design paid
attention to the feasibility of a concurrent operation of the legacy
systems as well as the precursor (Phase I) system. In this section,
we give a brief description of the field subsystems and communication
system of OWFA.

\begin{figure}
\centerline{ \includegraphics[width=\linewidth]{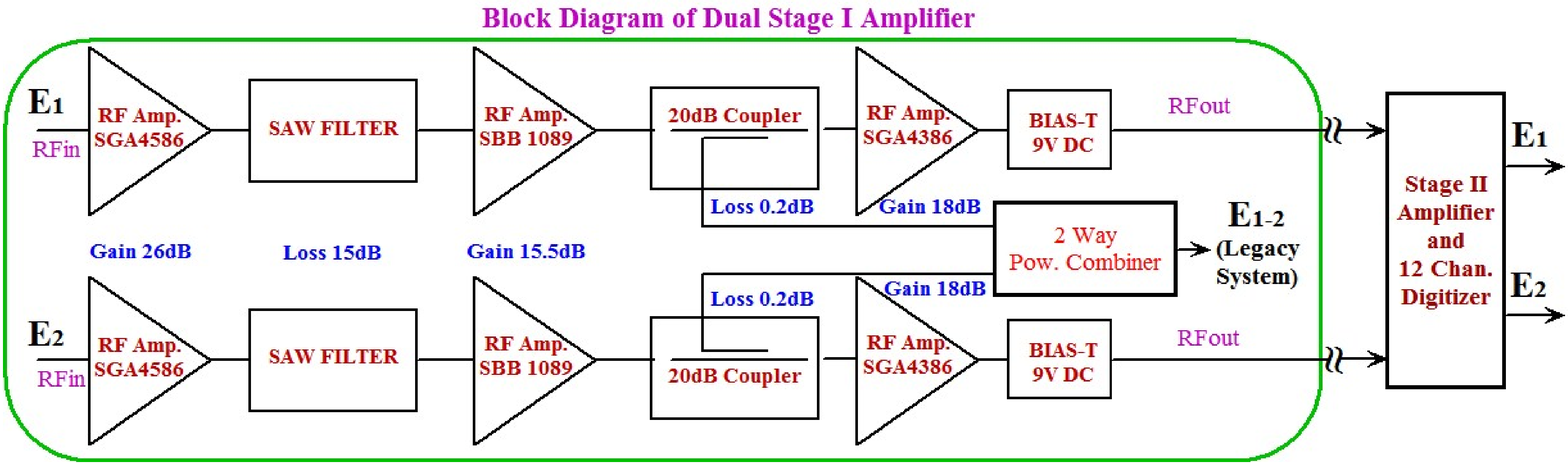}}
\caption{Dual Stage-1 Signal Conditioner. E1 and E2, respectively,
represent 1 to 4 dipoles and 5 to 8 dipoles combined outputs
(refer to Fig.~1).
}
\label{fig:Dual-Stage1}
\end{figure}

\subsection{Array Elements}

An antenna element of OWFA is defined as the phased sum of the
signals received from 4 consecutive dipoles along the focal line.
This corresponds to the output of a four-way combiner in the legacy
system (Figure \ref{fig:half-module}) equivalent to an antenna of size
30 m $\times$ 1.92 m for the dipole spacing of 0.48 m. In contrast with the
analog beam forming employed in the legacy system, the RF signals for OWFA
are directly digitized in the field. Signal conditioning for digitization is
done in two stages, with the first stage located inside the half-module
enclosure while the second stage located below the reflector in a new set of
metal enclosures (called “pillars”) installed close to the centre
of each module.

\begin{figure}
\centerline{\hskip 5in\includegraphics[height=4in]{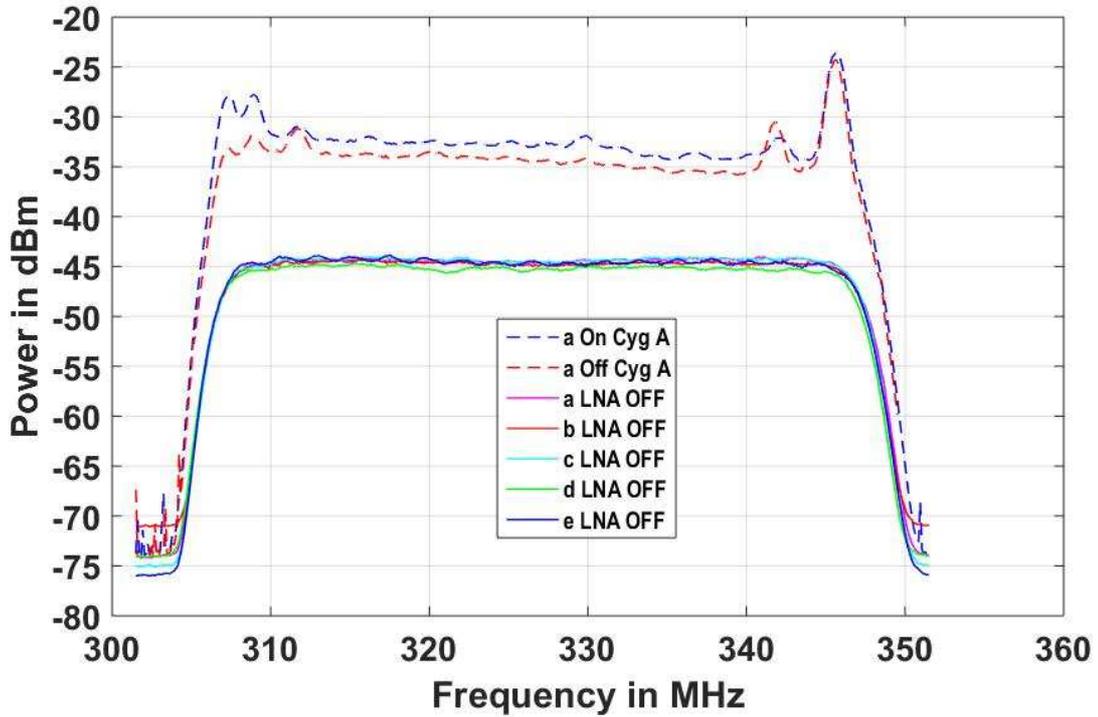}}
\caption{On source deflection for a single element. The long dash lines
  (ON Cygnus~A (blue curve), and OFF Cygnus~A (red curve)) show the total 
  power received from a single element, respectively, when pointed towards 
  Cygnus~A, and away 
  from Cygnus~A, showing the increase in antenna temperature when looking 
  at the bright source with a single 4-dipole element. The other continuous 
  lines (a LNA OFF, b LNA OFF, c LNA OFF, d LNA OFF, and e LNA OFF) are for 
  5 different elements for which the LNA has been terminated, {\it i.e.,} 
  they do not receive any sky signal. }
\label{fig:Bandpass}
\end{figure}

The first stage of signal conditioners consist of 3 units mounted
inside each half-module enclosure in place of two-way combiners of
the legacy system. As shown in Figure \ref{fig:Dual-Stage1}, each unit
consists of a pair of identical sub-units, each containing a multistage
amplifier bandlimited to about 38 MHz centred at 327.5 MHz using a commercially
available SAW filter. Coupled ports are provided from the individual
amplifiers whose outputs are combined to form a signal fully
compatible with that of the passive two-way combiner in the legacy
system. The normal outputs of the amplifiers are brought out on SMA
connectors to connect to a similar set of amplifiers located at the
base of the antenna. Thus, a total of six 60 m cables from each
half-module enclosure carry the RF to second stage signal conditioner.
The stage-2 amplifier is similar to the first stage (but with a
slightly higher gain) and includes a SAW filter identical to that
used in stage-1. A current injector in stage-2 and a diplexer in
stage-1 are used to transmit supply for stage-1 system through the
centre conductor of RF cable between the two stages. The output of
second stage signal conditioners are directly fed to the 12-channel
digitizer described below. The net gain of the system before
digitization is shown in Figure \ref{fig:Bandpass}. It may be recalled that the
bandpass is the result of cascading two SAW filters – one in stage-1
and the other in stage-2.

\subsection{Clock Distribution, Communication and control}

\begin{figure}
\centering{\includegraphics[width=\textwidth]{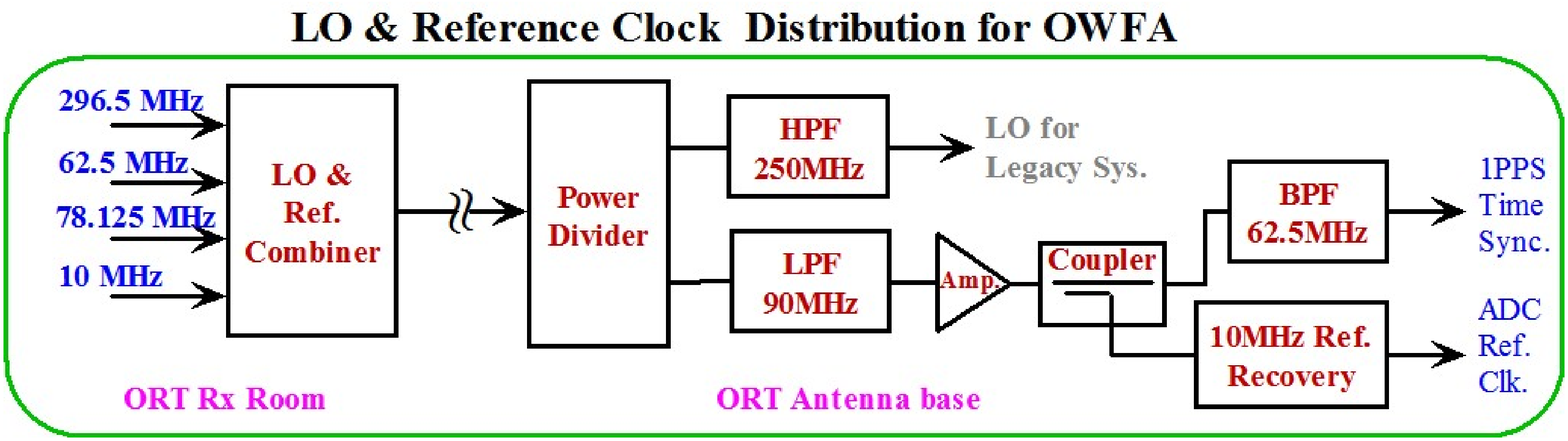}}
\caption{Reference and Sync Distribution}
\label{fig:RefDist}
\end{figure}

A GPS-disciplined rubidium oscillator provides a 10-MHz reference
which is distributed to all the modules along with the local
oscillator in the legacy system. A single cable originating from
the receiver room branches into 22 cables in a Christmas tree
configuration formed by a set of passive power dividers such that
the relative phase difference between various modules in the
transmission path is kept within a few degrees. At each module, a
high pass filter passes the 296.5 MHz to the legacy mixer unit and
a low pass filter passes frequencies below 100 MHz to OWFA subsystems
located below the reflector. A combination of frequencies below 100
MHz is generated and combined with the legacy local oscillator for
transmission through this chain (Figure \ref{fig:RefDist}). One of these frequencies
is the 10 MHz reference itself, from which the sampling clock is
locally synthesized in the 12-channel digitizer described below.
The other tones transmitted by OWFA system are meant for time
synchronization and phase monitoring.

Time synchronization between various modules is carried out using
a BFSK modulation with a phase-continuous frequency shift between
62.5 MHz and 78.125 MHz. The switching pattern is aligned to
one- pulse-per-second (1 pps) from a  GPS receiver at the transmission
end. For recovering the 1 pps at each module, BFSK is transformed
into an ASK modulation by passing the signal through a 62.5-MHz
filter to an rms detector (HMC10241).  The output of the rms detector
is converted to a sharp pulse using a Schmitt buffer and made
available to the digitizer and control cards in the corresponding
module as an external time sync.

\begin{figure}
\centering{\includegraphics[width=\linewidth]{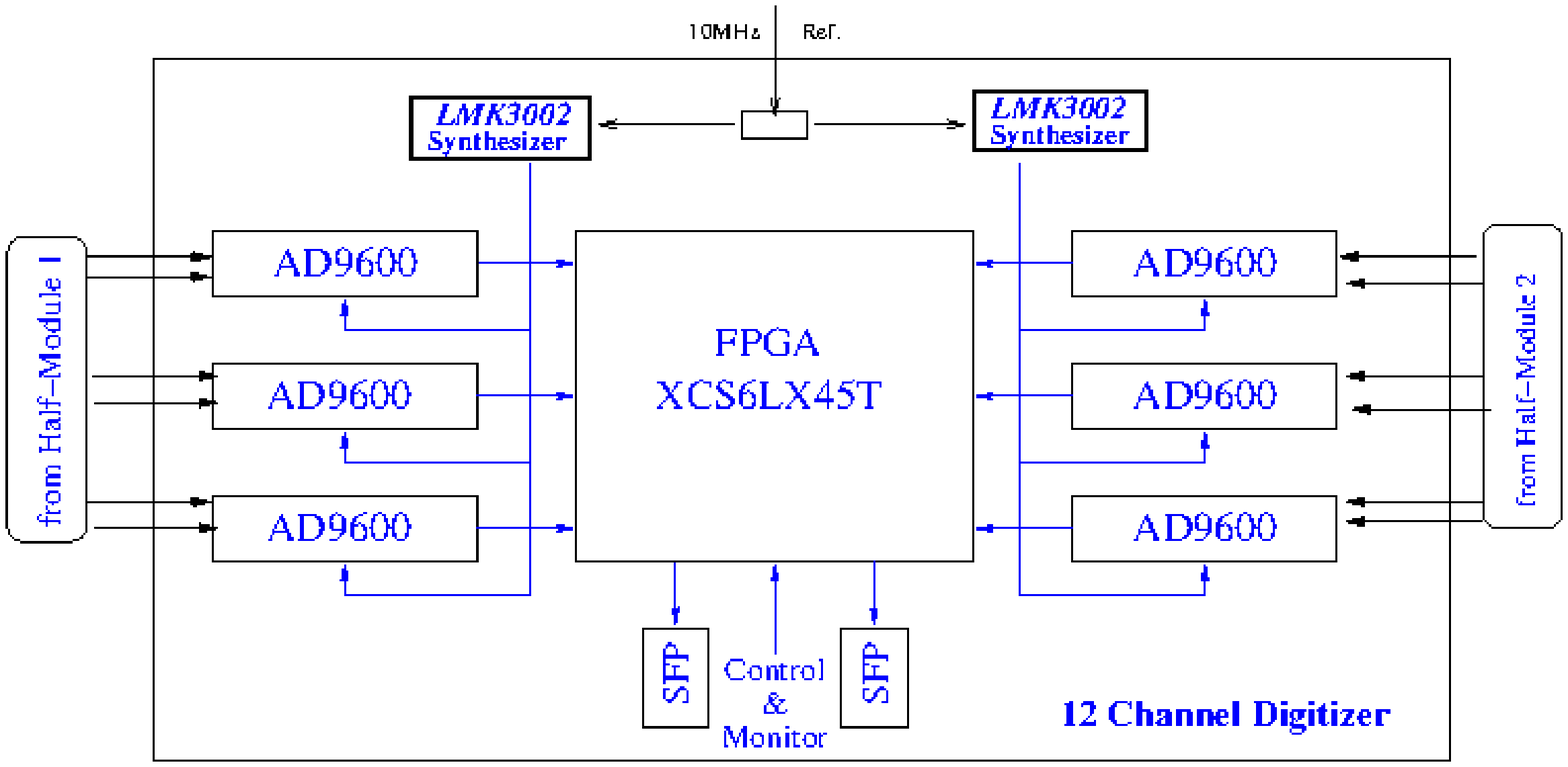}}
\caption{12-Channel Digitizer}
\label{fig:adc12}
\end{figure}

\subsection{12-Channel Digitizer}

As described earlier, at the base of each module, the RF signal elements from
the two corresponding half-modules (6 elements/half-module) are passed
through a second set of amplifiers (called stage-2 amplifiers) which
include an anti-aliasing filter identical to the SAW filter in
stage-1 amplifiers.  The output of these signal conditioners are
given to a 12-channel digitizer unit,  whose functional blocks are
shown in Figure \ref{fig:adc12}.  It consists of 6 dual-channel ADC devices
(AD9600) interfaced to an onboard FPGA (Spartan 6 SX45T).  A pair
of onboard synthesizers are used for generating the sampling clock
from the centrally distributed 10 MHz reference available at each
module.  In addition to providing the 76.8 MHz sampling clock to
the ADC devices,  a spare output from one
of the synthesizers is also connected to the on-board FPGA for its
housekeeping operations.  One of the house-keeping operations include
generation of local timer required for timestamping data. The local
timer generates a pulse per second (pps) used for internal logic.
During initialization, the internal pps is aligned with the time
sync recovered from the centrally distributed synchronization pulse.
A user command is also available to adjust the internal pps by
specified number of sampling clock ticks.  This also provides a
means for accommodating the small differences in propagation delays
from the central distribution point to the recovery points in various
modules.

From each ADC chip, the FPGA acquires data from the associated pair of
signals, and delay-compensates them using internal memory in the
device.  Delay-compensated data corresponding to 1600 consecutive
samples from the two signals connected to an ADC device are stored
in a memory block.  For this buffering, data are compressed using
a coding scheme which packs a total of 20 samples (10 from each
input signal) into a 64-bit word using a 3-bit coding and generating
4 control bits for each 64-bit word.  This results in a 1280B word
for 1600 consecutive samples from a pair of signals sampled by an
ADC device.  These are prefixed by a 32B preamble to form a frame
buffer of size 1312B.  Thus, at the end of 1600 sampling clock
ticks, a total of 6 frames of constant size (1312B) are formed,
where the 32B preamble in each frame contains  meta-data containing
useful information like sequence number, identification, timestamp,
delays and project code.  Double buffers are provided using on-chip
memory (Block Rams) so that while one set of 6 frames are being
filled in, the previous set of 6 frames are transported to the
central processing system.  For transportation,  the 6 frames are
handled as two sets of 3 frames, where each set contains data from
the 6 elements in a half-module.  These frames are serialized and clock-encoded
using Xilinx Aurora protocol and routed to two on-board transceivers
for transmission at 2.5 Gbps link speed. The physical layer is based
on standard off-the-shelf Small Form factor Plug-in (SFP) modules
appropriate for single mode fibres.

The important system parameters resulting from this digitization are
summarized in Table \ref{tab:SysPar}.

\begin{table}
\begin{center}
\caption{OWFA System Parameters}
\label{tab:SysPar}
\begin{tabular}[scale=0.5]{|l||l|}
\hline \hline Parameter      & OWFA                        \\
\hline\hline Band centre     & 326.5 MHz ($\lambda$ = 0.9182 m) \\
\hline Element size          & 1.92 m = $2.087\lambda$             \\
\hline Number of Elements    & 264                          \\
\hline Nominal FoV(NS)       & $27.5^{\circ}sec(\delta)$  \\
\hline Sampling Rate         & 76.8 Ms/s, 3-bit          \\
\hline Usable bandwidth      & $\sim$35 MHz typical     \\
\hline Continuum Sensitivity & 10 mJy$/\surd{t_{sec}} rms$  \\
\hline Spectral Resolution   & 48 {\rm kHz}              \\
\hline
\end{tabular}
\end{center}
\end{table}

\section{Conclusion}

Conversion of the legacy ORT to an interferometer (the ``Ooty Wide Field
Array'', OWFA) will result in an instrument with high sensitivity as
well as a large instantaneous field of view of $\sim$2$^\circ$ $\times$ 
28$^\circ sec(\delta)$. The OWFA will be equipped with an 800-channel FX
correlator with a spectral resolution of 48 kHz and a time resolution
of a few milliseconds.  This gives it a unique advantage for large
scale deep surveys at metre wavelengths. The main science drivers
are (1)~observations of the power spectrum of HI emission from large
scale structures at $z \sim 3.3$, (2) studies of the space weather in the
inner solar heliosphere and (3) studies of transient sources. More details on
the main science drivers behind the OWFA upgrade can be found in Subrahmanya,
Manoharan \& Chengalur (2016, this issue).

One of the major constraints in implementing this upgrade was the
requirement that the ORT be kept in constant use, and hence it is necessary
for the legacy receivers to continue to operate even as the upgrade
proceeds. This constraint was met by introducing amplifiers in the path
of passive combiner tree of the legacy system, from which one could
tap signals for both the legacy systems as well as for the OWFA
subsystems. If these amplifiers fail, the legacy system also will fail.
The front-end of OWFA has been switched on for almost two years without
affecting the legacy system.  Thus, the entire front-end subsystem of OWFA has
been undergoing field testing during this period and the successful
operation of the legacy system indicates their robustness and stability.

\section*{Acknowledgment}

We thank the members of the Radio Astronomy Laboratory at RRI (especially
S. Kasturi and S. Sujatha) for help during the initial prototyping and
design finalization. We acknowledge the enthusiastic support received from the
members of the Radio Astronomy Centre during the installation and commissioning
of the systems in the field. CRS and PKM wish to acknowledge the financial
support from ISRO for the project.

\bibliography{Receiverv2}{}
\bibliographystyle{plainnat}

\end{document}